\documentclass[11pt]{article}
\usepackage[pdftex]{graphicx}
\usepackage{hyperref}
\usepackage{footnote}
\makesavenoteenv{minipage}
\renewcommand{\title}[1]{%
    \bigskip%
    \begin{center}%
    \Large\bf #1%
    \end{center}%
    \vskip .2in}

\renewcommand{\author}[1]{%
    {\begin{center}
    #1
    \end{center}}}
\newcommand{\address}[1]{\vspace{-1.7em}\vspace{0pt}
    {\begin{center}
    \it #1
    \end{center}}}

\begin{document}

\begin{titlepage}
\title{Gauge symmetry and W-algebra in higher derivative systems}

\author{Rabin Banerjee $\,^{\rm a,b}$,
Pradip Mukherjee\footnote{Also, Visiting Associate, S. N. Bose National Centre 
for Basic Sciences, JD Block, Sector III, Salt Lake City, Kolkata -700 098, India  } $\,^{\rm c,d}$,
Biswajit Paul
$\,^{\rm a,e}$\ }
\address{$^{\rm a}$S. N. Bose National Centre 
for Basic Sciences, JD Block, Sector III, Salt Lake City, Kolkata -700 098, India }
\address{$^{\rm c}$Department of Physics, Presidency University,\\
86/1, College Street, Kolkata 700 073}
\address{$^{\rm b}$\tt rabin@bose.res.in}
\address{$^{\rm d}$\tt mukhpradip@gmail.com}
\address{$^{\rm e}$\tt bisu\_1729@bose.res.in}

\begin{abstract}
The problem of gauge symmetry in higher derivative Lagrangian systems is discussed from a Hamiltonian point of view. The number of independent gauge parameters is shown to be in general {\it{less}} than the number of independent primary first class constraints, thereby distinguishing it from conventional first order systems. Different  models have been considered as illustrative examples. In particular we show a direct connection between the gauge symmetry and the W-algebra for the rigid relativistic particle.
\end{abstract}

\end{titlepage}

%\noindent {\bf PAC codes:} 11.15.-q, 11.10.Nx,  \\
%{\bf{Keywords:}} 

%\end{titlepage}

\section{Introduction}
%%%%%%%%%%%%%%%%%%%%%%%%%%%%%%%%%%%%%%%%%%%%%%%%%%%%%%%%%%%%%%%%%%%%%%%%%%%%%%%%%%%%%%%%%%%%%%%%%%%
The concept of gauge symmetry occupies a pivotal place in modern theoretical physics. The Standard model of elementary particle physics is entrenched on the gauge principle. Gravity/gauge duality is gaining  increasing importance. No wonder that a significant effort is continually being spent on the analysis of gauge symmetries. In this context the canonical methods constitute a very important part. Of these approaches a central role is played by Dirac's theory of constrained Hamiltonian dynamics \cite{D}. It provides a basic tool for understanding gauge symmetry of classical Lagrangian systems and forms the starting point of their quantization.
% The first class constraints of the theory generate gauge transformations and the number of independent primary first class constraints give the number of independent gauge degrees of freedom.
More than half a century of intense research in this field \cite{D, Sudarshan, HRT,Sunder,rothe} has gone a long way towards the identification, categorization and classification of the gauge invariances in their most general form. It is notable in this context that the vast resources of Hamiltonian analysis of gauge invariances available in the literature\cite{Komar, Teitelboim,C,CGS,  HTZ, Pons1,  BRR, BRR1, BRR2, Pons2, MS, BGMR} are almost confined within the realm of  usual Lagrangian systems where the Lagrangian is a function of coordinates and velocities only. The important sector of the higher order Lagrangians remains considerably less explored. Of course much work has been done on the Hamiltonian analysis of higher derivative Lagrangian systems \cite{Ostro, BGPR, pl1,  N,  B, Mo, DS, gitman2, AGMM}.
% There are seminal works on the subject that explore various aspects of such theories.
 However some important issues are pending to be addressed. One such issue is the abstraction of the independent gauge transformations from the first class constraints of a gauge theory, as usually done in  conventional systems. For  usual systems the first class constraints are shown to generate gauge transformations (GT). However, not all such GTs are independent. The number of independent GTs has been shown to be equal to the number of independent primary first class constraints (PFC). There are very definite algorithms \cite{C, HTZ, BRR, BRR1, BRR2} of finding the  gauge generator 
containing just the right number of independent gauge parameters. Similar analysis, however, is lacking for  higher derivative systems.

  Lagrangian theories with higher order derivatives are interesting in their own right. These have been discussed over a long period of time
\cite{lw,gitman1, gitman2,P, pl1,  N,  DN, HH, R, S, K, cl, AGMM, M}.
Such theories have appeared in different contexts: higher derivative terms naturally occur as quantum corrections to the lower order theories. Higher-derivative Lagrangians appear to be a useful tool to describe some interesting models like relativistic particles with rigidity, curvature and torsion \cite{P, pl1}. Various stringy models are shown to be equivalent to higher derivative theories.  Adding higher derivatives may improve
ultraviolet behaviour \cite{1b,2b}. They make modified
gravity renormalizable \cite{3b} or even asymptotically free \cite{4b}. The literature is rich in possible higher derivative theories of gravity which find application in quantum gravity \cite{BOS}. More recently, the higher derivative systems appear in $f(R)$ gravity, which is of cardinal significance in explaining new evidences coming from astrophysics and cosmology such as the late time acceleration of the universe \cite{Sot}. Understanding the gauge invariances of higher derivative Lagrangian systems is thus essential in the context of current theoretical physics research. 
%We find that although the methods developed for 
% systems are applicable there are nontrivial differences. Also, the role of dispersion relations vis-a-vis the algebra of coordinates get highlighted.

%  The issue of the gauge symmetry of the higher derivative theories has characteristic features that differentiate it from usual first order theories. It has shown to lead to hitherto unexpected connections. Thus the gauge symmetry of the scale -- invariant rigid particle model is demonstrated to be associated with the $W_3$ symmetry of Zamolodchikov's W -- algebras \cite{RR}. On the other hand the problem has peculiar surprises. In connection with relativistic particle model with action solely determined by curvature, Hamiltonian gauge generators are reported to vanish, though Lagrangian gauge generators were obtained \cite{GP}. The importance of a systematic study of the gauge invariances of the higher derivative theories thus cannot be overemphasised.
 
  In the present paper we would like to address the above mentioned problem. Specifically we want to explore what new restrictions
are imposed on the gauge generator by the phase space structure peculiar to the higher derivative theories. For  conventional first order systems there exists a Hamiltonian method  which treats the gauge symmetries as transformations between different field configurations without taking recourse to the equations of motion \cite{BRR}. This analysis has been applied to numerous models in the literature \cite{BMS1, BMS2, MS, GGS, BGMR}. We would like to adapt this method to  higher order theories. At  first sight it appears that the number of independent gauge transformations will naturally be equal to the number of independent PFCs, exactly as happens for  first order systems.
%even in case of the higher derivative theories as our Hamiltonian formalism is based on an equivalent first order formalism.
 But we will see that this is not true and in general  the number of independent gauge transformations may be {\it{less}} than the number of PFCs.
%in the process of which we will see that there are surprises inherent in the process.
The reason behind this is the peculiarities inherent in  higher derivative theories. We have elaborated this characteristic feature of  higher derivative theories by taking up the examples of different relativistic particle models  \cite{P} which offer a simple setting for the study of higher derivative systems. 
  
  The relativistic particle in its standard version where the action is given by the integrated proper time, is a popular model because it offers a simple realization of diffeomorphism invariant system. It is easy to check that the action is invariant under reparametrization of the particle trajectory in space -- time. This reparmetrization invariance is shared by more pertinent theories such as string theory and gravity.  More general particle models with the action modified by curvature and/or torsion terms have been investigated for a long time with many important results. Now the relativistic particle models with curvature and torsion dependent terms in the action \cite{P} contain higher derivative terms in the Lagrangean and thus constitute an important class of higher derivative theories. This higher derivative feature immediately contributes characteristic peculiarities. In the Hamiltonian analysis of first order systems the reparametrization symmetry manifests itself as gauge symmetry and parameters of one set of transformations can be exactly mapped to the parameters of the other set \cite{BMS1, BMS2, MS, GGS}. In particular this feature is revealed by the ususl relativistic particle model which is a first order theory. 
%The relativistic particle models with curvature and torsion dependent terms in the action \cite{P,pl2, Pl11} constitute an important class of higher derivative theories. These models are g In the Hamiltonian analysis the reparametrization symmetry manifests itself as gauge symmetry and parameters of one set of transformations can be exactly mapped to the parameters of the other set \cite{BMS1, BMS2, MS, GGS}. 
 A violation of this occurs in the higher derivative relativistic particle models with curvature term in the action ( massive relativistic particle with rigidity )\cite{N}. 
  % which is not even noticed much less emphasised. Thus for the relativistic particle with curvature treated in \cite{N} 
 Here there are two PFCs so that the reported gauge generator apparently consists of two independent gauge parameters. On the other hand, there is only one independent reparametrization parameter. It appears that the correspondence between gauge and reparametrization symmetry is lost, a fact which is not even noticed much less emphasised. The solution of this paradox is clearly our finding stated in the earlier paragraph --  the number of independent gauge parameters is  in general {\it{less}} than the number of primary first class constraints. We thus take up the model of \cite{N} and by applying our general method establish a one to one mapping between the gauge and the reparametrization parameters. In the course of our analysis we also provide a formally new Hamiltonian analysis of the model.

 The issue of the gauge symmetry of the higher derivative relativistic particle models  indeed has characteristic features that differentiate it from usual first order theories. It has also shown to lead to hitherto unexpected connections. Thus the gauge symmetry of the scale -- invariant rigid particle model, where the action is solely given by the curvature term, is demonstrated to be associated with the $W_3$ symmetry of Zamolodchikov's W -- algebras \cite{RR}.  The absence of the mass term introduces this new W -- symmetry in addition to the well known diffeomorphism invariance. It is notable that this extra symmetry is revealed by casting the equations of motion in the Boussinesq form. By examining the most general transformations that preserve the structure of the Boussinesq Lax operators it was  demonstrated that the symmetry group  of the model satisfies $W_3$ algebra \cite{RR}. A consistent hamiltonian analysis of the model which will reveal the full symmetries will thus be very welcome. We thus take up the relativistic particle model with action solely determined by curvature ( the rigid relativistic particle ) as our second illustrative example. As will be detailled in the following, our general approach successfully exhibits the full symmetry group, including both diffomorphism and W-morphism. This is important in view of the fact that previous analysis of gauge symmetries of the model reported the vanishing of hamiltonian gauge generators, though Lagrangian gauge generators were obtained \cite{GP}. 
 %Indeed, by explicitly computing the gauge transformation of the basic fields from the Hamiltonian gauge generator we show the invariance of the action. In other words the w-symmetry found here is an invariance of the action.A similar finding was reported earlier which however was based on an on-shell tratment requiring the construction of the appropriate Lax operators.
% The importance of a systematic study of the gauge invariances of the higher derivative theories thus cannot be overemphasised.       

 Before proceeding further let us describe the organisation of the paper. As already stated we will perform the Hamiltonian analysis in the framework of an equivalent first order formalism.
% \cite{GR}.
% In our study of the gauge symmetry of the higher derivative theories the Hamiltonian algorithm of \cite{BRR} will thus be very useful. Hence in section 2 a brief review of this   algorithm is presented. In the following section gauge invariance of 
We have demonstrated in section 2 that for the higher derivative theories 
%is discussed from a general setting where we argue why for such theories 
the  gauge generator should
 in general, contain lesser number of parameters than the number of independent PFCs of the theory. Note that this mismatch led to a confusion in an earlier calculation of relativistic particle model with curvature \cite{N}
% on a specific higher derivative model, 
which, however, was not emphasised. % The model of \cite{N} is that of the relativistic particle with curvature \cite{P, Pl2, pl1}. 
This model is invariant under the reparametrization of particle trajectory. The infinitesimal reparametrization consists of one independent reparametrization parameter. This reparametrization should be equivalent to the transformations generated by the Hamiltonian generator which, however, apparently contain two independent gauge parameters. In the following section we have presented a new Hamiltonian analysis of the model and use the general formalism of section 2 to resolve the apparent inconsistency. We also provide a mapping between the gauge and the reparametrization parameters. Our method proposes a new algorithm to find out the gauge invariances of higher derivative theories. It is remarkable that it offers  a consistent mapping between the gauge and reparametrization parameters, hitherto unavailable in the literature. 
%The internal consistency of our algorithm is further checked in section 3.

   Our analysis in section 3.1 shows that the particle momentum is unconstrained and the standard dispersion relation does not follow as a constraint as in the usual relativistic particle model. On the other hand this condition appears as a singular point in our analysis. The analysis of section 3.1 is valid excluding this singular point. However, imposition of this singular condition is interesting in its own right. The constraint structure is modified non-trivially leading a new symplectic algebra. We have studied the behaviour at the singular point in section 3.2. Finally the internal consistency of our algorithm is checked in section 3.3.

 After considering the relativistic massive particle with rigidity we take up the rigid relativistic particle model. The action of the later is obtained by putting the mass term zero in the model considered in section 3 but the symmetry structure is radically different as we have discussed in the above. In section 4.1 we give the Hamiltonian analysis of the model. In section 4.2 the formalism of section 2 is applied which leads to the realization of $W_3$ algebra.
The paper ends with a few concluding remarks in section 5.
% Finally a consistency check of our formalism based on (\ref{218})is provided in the appendix.
%Illustration of our method of analysing the gauge symmetry of the higher derivative theories is presented in section 4 taking up the model of \cite{N} as an example. A new Hamiltonian analysis of the model is given along with a resolution of the apparent mismatch between the independent number of gauge and rerparametrization parameters. In section 5 we have checked the consistency of our analysis.
% Our concluding remarks are contained in section 6.

\section{ General formalism}
%\section{Gauge symmetries of  higher order Lagrangian systems}

   We begin with a general higher derivative theory given by the Lagrangian
\begin{equation}
L = L\left(x, \dot{x}, \ddot{x}, \cdots , x^{\left(\nu\right)}\right)
\label{originallagrangean}
\end{equation}
where $x = x_n(n = 1,2,\cdots,\nu)$ are the coordinates and $\dot{}$ means derivative with respect to time. $\nu$-th order derivative of time is denoted by $x^{\left(\nu\right)}$. %For convenience of discussion we consider a second order theory
%\begin{equation}
%L = L\left(x, \dot{x}, \ddot{x}\right)
%\end{equation}
The Hamiltonian formulation of the theory may be conveniently done by a variant of Ostrogradskii method. The crux of the method consists in embedding  the original higher derivative theory to an effective first order theory. We define the variables $q_{n,\alpha} \left(\alpha = 1, 2, ...., \nu - 1 \right)$ as
\begin{eqnarray}
q_{n,1}   &=& x_n\nonumber\\
q_{n,\alpha} &=& \dot{q}_{n,\alpha -1}, \left(\alpha > 1 \right)
\label{newvariables}
\end{eqnarray}
This leads to the following Lagrangian constraints
\begin{eqnarray}
q_{n,\alpha} - \dot{q}_{n,\alpha -1} = 0, \left(\alpha > 1 \right)
\label{lagrangeanconstraints}
\end{eqnarray}
which must be enforced by corresponding Lagrange multipliers .   
The auxiliary Lagrange function of this extended description
of the system is given by
\begin{eqnarray}
L^*(q_{n,\alpha},\dot{q}_{n,\alpha},\lambda_{n,\beta})
=L\left(q_{n,1},q_{n,2}\cdots,q_{n,\nu-1},
\dot{q}_{n,\nu-1}\right)+_n\sum_{\beta=2}^{\nu-1}
\left(q_{n,\beta}-\dot{q}_{n,\beta-1}\right)\lambda_{n,\beta}\ ,
\label{extendedlagrangean}
\end{eqnarray}
where $\lambda_{n,\beta} (\beta = 2,\cdots , \nu - 1)$ are the Lagrange multipliers. If we consider these multipliers as independent fields then the Lagrangian $L^*$ becomes first order to which the well known methods of Hamiltonian analysis for  first order systems apply.
The momenta canonically
conjugate to the degrees of freedom $q_{n,\alpha}$,
$(\alpha=1,2,\cdots,\nu-1)$ and
$\lambda_{n,\beta}$ $(\beta = 2, \cdots,\nu-1)$
are defined, respectively, by,
\begin{equation}
p_{n,\alpha}=\frac{\partial L^{*}}{\partial \dot{q}_{n,\alpha}}\ ,\ \
\pi_{n,\beta}=\frac{\partial L^{*}}{\partial\dot{\lambda}_{n,\beta}}\ .
\end{equation}
These immediately lead at least to the following 
primary constraints,
\begin{equation}
\Phi_{n,\beta} \approx 0\ ,
\ \ \pi_{n,\beta} \approx 0\ ,\ \
\beta = 2,\cdots,\nu -1\ ,
\label{constraints}
\end{equation}
where
\begin{equation}
\Phi_{n,\beta}\equiv p_{n,\beta -1}+\lambda_{n,\beta}\ , \ \
\beta = 2,\cdots,\nu - 1\ .
\end{equation}
Note that depending on the situation whether the original Lagrangian $L$ is singular there may be more primary constraints.\\ 
Let us first assume  that the original Lagrangian $L$ is regular. Then there are no more primary constraints. Now
the basic non-trivial Poisson brackets are
\begin{equation}
\left\{q_{n,\alpha},p_{m,\alpha '}\right\}=\delta_{nm}\delta_{\alpha \alpha '}\ ,\ \
\left\{\lambda_{n,\beta},\pi_{m,\beta '}\right\}=\delta_{nm}\delta_{\beta \beta '}
\ ,
\label{basicpbs}
\end{equation}
Consequently the primary constraints obey the algebra 
\begin{equation}
\left\{\Phi_{n,\beta},\Phi_{m,\beta '}\right\}=0\ ,\ \
\left\{\Phi_{n,\beta},\pi_{m,\beta '}\right\}=
\delta_{nm}\delta_{\beta \beta '}\ ,\ \
\left\{\pi_{n,\beta},\pi_{m,\beta '}\right\}=0\ ,
\label{algebraconstraints}
\end{equation}
implying that $\pi_{n,\beta}$ and $\Phi_{n,\beta}$ are second class constraints. Now, the 
ca\-no\-ni\-cal Hamiltonian of the modified system(\ref{extendedlagrangean}) can be written according to the usual prescription as,
\begin{equation}
H_C=\sum_n\sum_{\alpha=1}^{\nu-1}\dot{q}_{n,\alpha}p_{n,\alpha}
+\sum_n\sum_{\beta=2}^{\nu-1}\dot{\lambda}_{n,\beta}\pi_{n,\beta}-L^*\ ,
\end{equation}
We define the total Hamiltonian ($H_T$) by adding linear combinations of the primary constraints (\ref{constraints}), 
\begin{equation}
H_T = H_C + u_{n,\beta}\pi_{n,\beta} + v_{n,\beta}\Phi_{n,\beta}
\end{equation}
where $ u_{n,\beta}$ and $v_{n,\beta}$ are Lagrange multipliers. From (\ref{algebraconstraints}) we found that the constraints are second class. Thus preserving the primary constraints in time we will be able to fix the multipliers $ u_{n,\beta}$ and $v_{n,\beta}$. Fixing these and after some simplifications we find that
\begin{equation}
H_T\left(q_{n,\alpha},p_{n,\alpha};\lambda_{n,\beta}\right)=
\overline{H}_0\left(q_{n,\alpha},p_{n,\nu-1}\right)
-\sum_n\sum_{\beta=2}^{\nu-1}\lambda_{n,\beta}q_{n,\beta}\ .
\label{eq:concanH0}
\end{equation}
\begin{equation}
\overline{H}_0(q_{n,\alpha},p_{n,\nu-1})=
\sum_n\dot{q}_{n,\nu-1}p_{n,\nu-1}-
L\left(q_{n,\alpha},\dot{q}_{n,\nu-1}\right)\ .
\label{eq:H0restricted}
\end{equation}
The presence of the  second class constraints (\ref{constraints}) imply that the phase space degrees of freedom
$(q_{n,\alpha},p_{n,\alpha};\lambda_{n,\beta},\pi_{n,\beta})$ are
not all independent. If we replace the Poisson brackets by Dirac brackets the constraints (\ref{constraints}) can be strongly implemented. This enables us to eliminate the nondynamical sector $(\lambda_{n,\beta},\pi_{n,\beta})$ from the phase space variables. Straightforward calculations show that the DBs between the remaining phase space variables are the same as the corresponding PBs. The total Hamiltonian now becomes
\begin{equation}
H_T\left(q_{n,\alpha},p_{n,\alpha};\lambda_{n,\beta}\right)=
\overline{H}_0\left(q_{n,\alpha},p_{n,\nu-1}\right)
+ \sum_n\sum_{\beta = 2}^{\nu-1}p_{n,\beta - 1}q_{n,\beta}\ .
\label{eq:canH0}
\end{equation}
We can go on writing the canonical equations and so on. The Hamiltonian analysis of the regular theory is complete.

         The interesting case for us is when $L$ is singular. Now the following possibilities may arise:
\begin{enumerate}

\item The original Lagrangian is singular but the additional constraints are all second class. Conserving the full set of primary constraints in time does not yield any secondary constraint. Rather,  all the multipliers in the total Hamiltonian will get fixed. The reduction of phase space may be done by implementing the second class constraints strongly provided we replace all the PBs by appropriate DBs.  

\item The original Lagrangian is singular and there are both primary second class and first class constraints among them. Conserving the primary constraints in time, secondary constraints will now be obtained. There may be both secondary second class and first class constraints. The second class constraints may be eliminated again by the DB technique. The first class constraints generate gauge transformations which are required to be further analysed. These constraints may yield further constraints and so on. The iterative process stops when no new constraints are generated. 
\end{enumerate}
From the point of view of gauge invariance the second case is important. Since the original Lagrangian system is replaced by the first order theory (\ref{extendedlagrangean}) the algorithm of \cite{BRR, BRR1} can be readily applied. All the first class constraints appear in the gauge generator G
%(see equation (\ref{217})):
\begin{equation}
G = \sum_a \epsilon^a \Phi_a
\label{217}
\end{equation}
where $\{\Phi_a\}$ is the whole set of (primary and secondary) first class constraints and $\epsilon^a$ are the gauge parameters. These parameters are however not independent. For a first order system the number of independent gauge parameters is equal to the number of independent PFCs. % For the first order Lagrangians, the requirement of the commutativity of the time derivative and the gauge variation can be exploited to express the gauge generator in terms of a set of gauge parameters whose number equals the number of primary first class constraints. 
Following the algorithm of \cite{BRR, BRR1} we can express the dependent gauge parameters in terms of the independent set using the conditions
%\begin{equation}
%\delta\lambda^{a_1} = \frac{d\epsilon^{a_1}}{dt}
%                 -\epsilon^{a}\left(V_{a}^{a_1}
%                 +\lambda^{b_1}C_{b_1a}^{a_1}\right)
%                             \label{218}
%\end{equation}
\begin{equation}
  0 = \frac{d\epsilon^{a_2}}{dt}
 -\epsilon^{a}\left(V_{a}^{a_2}
+\lambda^{b_1}C_{b_1a}^{a_2}\right)
\label{219}
\end{equation}
The indices $a_1, b_1 ...$ refer to the primary first class constraints while the indices $a_2, b_2 ...$ correspond to the secondary first class constraints.
The coefficients $V_{a}^{a_{1}}$ and $C_{b_1a}^{a_1}$ are the structure
functions of the involutive algebra, defined as
\begin{eqnarray}
\{H_c,\Phi_{a}\} = V_{a}^b\Phi_{b}\nonumber\\
\{\Phi_{a},\Phi_{b}\} = C_{ab}^{c}\Phi_{c}
\label{2110}
\end{eqnarray}
and $\lambda^{a_1}$ are the Lagrange multipliers(associated with the primary first class constraints)  appearing in the expression of the total Hamiltonian.
Solving (\ref{219}) it is possible to choose $a_1$ independent
gauge parameters from the set $\epsilon^{a}$ and express $G$ of
(\ref{217}) entirely in terms of them. For the conventional first order theories this completes the picture. 
The situation for higher order theories is, however, different.
 %However unlike the first order theories this set of parameters will not be mutually independent in general.
%For the higher order theories the number of independent gauge parameters will be even less.
 This is because of the new constraints (\ref{lagrangeanconstraints}) appearing in the effective first order lagrangean (\ref{extendedlagrangean}). Owing to these  we additionally require
\begin{eqnarray}
\delta q_{n,\alpha} - \frac{d}{dt}\delta{q}_{n,\alpha -1} = 0, \left(\alpha > 1 \right)
\label{varsgauge}
\end{eqnarray}
%where $\delta A$ is the gauge variation of the phase space variable $A$.
%obtained from (\ref{lagrangeanconstraints}).
% by using (\ref{commutativity1}).
 These conditions may  reduce the number of independent gauge parameters further. Thus the number of independent gauge parameters is, in general, less than the number of primary first class constraints. This shows the difference from a genuine first order theory.
 % The conversion of the original higher derivative theory to an effective first order theory is thus a formal one containing peculiarities characteristic of higher derivative nature. 
 
 The effective first order theory, obtained from the original higher order theory, therefore contains peculiarities. In the following we will illustrate this peculiarity using the examples of different relativistic point particle models containing higher derivative terms.

%Before going to the applications of the general formalism a few words about the consistency of the formalism will be appropriate. 
%Fortunately, in the algorithm of \cite{BRR, BRR1} there is a provision of checking the consistency of calculation. 
%The gauge vatriation of the Lagrange multipliers satisfy
%\begin{equation}
%\delta\lambda^{a_1} = \frac{d\epsilon^{a_1}}{dt}
%                 -\epsilon^{a}\left(V_{a}^{a_1}
%                 +\lambda^{b_1}C_{b_1a}^{a_1}\right)
%                             \label{218}
%\end{equation}
%It can be shown that the condition (\ref{218}) actually follows from (\ref{219}) \cite{BRR,BRR1}. Also the same gauge variations can be directly obtained as $$\delta\lambda^{a_1} = \{\lambda^{a_1}, G\}$$ where we assume that the generator $G$ is expressed in terms of the independent gauge parameters. If $\delta\lambda^{a_1}$ The agreement of both the expressions of $\delta\lambda^{a_1}$ will provide a nontrivial consistency check of our analysis.
 
\section{Relativistic particle model with curvature}
	The Hamiltonian formulation is developed with the aim of constructing the gauge generator and abstracting the symmetries of the model. In the course of this analysis we show that the nature of the constraints is non-trivially modified at the critical point $p_{1}^{2} = m^{2}$ that corresponds to the standard dispersion relation. Hence the analysis at this point is separately given. 
\subsection{Hamiltonian analysis and construction of the gauge generator}
The massive relativistic point particle theory with rigidity \cite{P}  has the action\footnote{contractions are abbreviated as $A^\mu B_\mu = AB$, $A^\mu A_\mu=A^2$. We consider the model in 3 + 1 dimensions. So $\mu$ assumes the values 0, 1, 2, 3. Also, the model is meaningful for  $\alpha < 0 $ \cite{pl1}} 
\begin{equation}
S=-m \int{\sqrt{\dot{x}^{2}}}d\tau+\alpha\int{\frac{\left( \left( \dot{x}\ddot{x}\right)^{2}-\dot{x}^{2}\ddot{x}^{2} \right)^\frac{1}{2}}{\dot{x}^{2}}}d\tau\label{maction}
\end{equation}
We introduce the new coordinates
\begin{equation}
q^{\mu}_1 = x^{\mu}\ \;\ \ q^{\mu}_2 = \dot{x}^{\mu}
\label{mnewcoordinate}
\end{equation}
The Lagrangian in these coordinates has a first order form given by \cite{pl1}
\begin{eqnarray}
% \nonumber
%L_{0} &=& -m\sqrt{q_{2}^{2}} + \alpha \frac{\left({\left({q_{2}\dot{q}_{2}} \right)^{2} - q_{2}^{2} \dot{q}_{2}^{2} } \right)^{\frac{1}{2}} }{q_{2}^{2}}
%\nonumber \\
L &=& -m\sqrt{q_{2}^{2}} + \alpha \frac{\left({\left({q_{2}\dot{q}_{2}} \right)^{2} - q_{2}^{2} \dot{q}_{2}^{2} } \right)^{\frac{1}{2}} }{q_{2}^{2}} + q_{0}^{\mu}(\dot{q}_{1\mu} - q_{2\mu})
\label{mindependentfields}
\end{eqnarray}
where $q_{0}^{\mu}$ are the Lagrange multipliers that enforce the constraints
\begin{equation}
\dot{q}_{1\mu} - q_{2\mu} = 0
\label{27}
\end{equation}
Let $p_{0\mu}$, $p_{1\mu}$ and $p_{2\mu}$ be the canonical momenta conjugate to $q_{0\mu}$, $q_{1\mu}$ and $q_{2\mu}$ respectively. 
Then 
we immediately get the following
primary constraints
\begin{eqnarray}
%\nonumber
\Phi_{0\mu} = p_{0\mu} \approx 0 ;
\Phi_{1\mu} = p_{1\mu} - q_{0\mu} \approx 0
\label{mPFC1}
\end{eqnarray}
and
\begin{eqnarray}
%\nonumber
\Phi_{1} = p_{2}q_{2} \approx 0;
\Phi_{2} = p_{2}^{2}q_{2}^{2} + \alpha^{2} \approx 0
\label{mPFC2}
\end{eqnarray}
The first set of constraints (\ref{mPFC1}) are an outcome of our extension of the original Lagrangian. The canonical Hamiltonian following from the usual definition is given by
\begin{equation}
H_{C} =  m\sqrt{q_{2}^{2}} + q_{0\mu}q_{2}^{\mu}
 \label{CH}
\end{equation}
The total Hamiltonian is
\begin{equation}
H_{T} =  H_{C} + u_{0\mu}\Phi_{0}^{\mu} + u_{1\mu}\Phi_{1}^{\mu} + \xi^{1}\Phi_{1} + \xi^{2}\Phi_{2} 
\end{equation}
where $u_{0\mu}$, $u_{1\mu}$,$\xi^{1}$ and $\xi^{2}$ are as yet undetermined multipliers. Now, conserving the primary constraints
% we get
%\begin{eqnarray}
%\nonumber 
%\dot{\Phi}_{0\mu} &=& \{\Phi_{0\mu}, H_T\}  = -q_{2\mu} + u_{1\mu} \approx 0
%\nonumber \\
%\dot{\Phi}_{1\mu} &=& \{\Phi_{1\mu}, H_T\}  = -u_{0\mu} \approx 0
%\nonumber \\
%\dot{\Phi}_{1} &=& \{\Phi_{1},  H_T\}  = -(q_{0}q_{2} + m\sqrt{q_{2}^{2}}) \approx 0
%\nonumber \\
%\dot{\Phi}_{2} &=& \{\Phi_{2},  H_T\} = 2(q_{0}p_{2})q_{2}^{2} \approx 0
%\label{34}
%\end{eqnarray}
we find that the multipliers  $u_{0\mu}$ and $u_{1\mu}$ are fixed:
\begin{equation}
u_{0\mu} = 0 \ \ and \ \ u_{1\mu} = q_{2\mu}
\end{equation}
Also new
secondary constraints are obtained
\begin{eqnarray}
%\nonumber
\omega_{1} =  q_{0}q_{2} +  m\sqrt{q_{2}^{2}} \approx 0;
\omega_{2} = q_{0}p_{2} \approx 0
\label{mssc}
\end{eqnarray}
The last constraint in (\ref{mssc}) is obtained by assuming $q_{2}^{2} \neq 0$ which follows from
 the structure of the Lagrangian (\ref{mindependentfields}). The total Hamiltonian now becomes
\begin{equation}
H_{T} = m\sqrt{q_{2}^{2}} + q_{0\mu}q_{2}^{\mu} + q_{2\mu}\Phi_{1}^{\mu} + \xi^{1}\Phi_{1} + \xi^{2}\Phi_{2}
\end{equation}
PBs among the constraints are given by \\
\begin{eqnarray}
\nonumber
\left\lbrace {\Phi_{0\mu}, \omega_{1}}\right\rbrace &=& -q_{2\mu}
\nonumber \\
\left\lbrace {\Phi_{0\mu}, \omega_{2}}\right\rbrace &=& -p_{2\mu}
\nonumber \\
\left\lbrace {\Phi_{1}, \omega_{1}}\right\rbrace &=& -\omega_{1} =0
\nonumber  \\
\left\lbrace {\Phi_{1}, \omega_{2}}\right\rbrace &=& \omega_{2} = 0
\nonumber \\
\left\lbrace {\Phi_{2}, \omega_{1}}\right\rbrace &=& -2( \omega_{2} + m\sqrt{q_{2}^{2}}\Phi_{2} ) = 0
\nonumber \\
\left\lbrace {\Phi_{2}, \omega_{2}}\right\rbrace &=& 2p_{2}^{2}( q_{0}q_{2} )
\nonumber \\
\left\lbrace{\omega_{1}, \omega_{2}} \right\rbrace &=& q_{0}^{2} - m^{2}
\label{consalgebra}       
\end{eqnarray}
 Conserving the secondary constraints we get
%find that 
%no tertiary constraints are obtained but
\begin{eqnarray}
\nonumber
\dot{\omega}_{1} &=& 0
\nonumber \\
\dot{\omega}_{2}  &=& -q_{0\mu}\left(  {\frac{m}{\sqrt{q_{2}^{2}}}q_{2}^{\mu} + q_{0}^{\mu}} \right) - 2\xi^{2} p_{2}^{2}(q_{0}q_{2}) = 0\label{condition1}
\end{eqnarray}
Clearly no more tertiary constraints are obtained. 
The second condition of (\ref{condition1}) fixes 
the multiplier $\xi^{2}$ 
%is fixed, so that the iterative process terminates;\\
\begin{equation}
\xi^{2} = - \frac{1}{2p_{2}^{2}(q_{0}q_{2})}\left(  {\frac{m}{\sqrt{q_{2}^{2}}}(q_{0}q_{2}) + q_{0}^{2} } \right) 
\end{equation}
Substituting this in the expression of the total Hamiltonian we get
\begin{eqnarray}
%\nonumber
H_{T} =
%  m\sqrt{q_{2}^{2}} + q_{0\mu}q_{2}^{\mu} + q_{2\mu}\Phi_{1}^{\mu} + \xi_{1}\Phi_{1}  - \frac{1}{2p_{2}^{2}(q_{0}q_{2})}\left(  {\frac{m}{\sqrt{q_{2}^{2}}}(q_{0}q_{2}) + q_{0}^{2} } \right)\Phi_{2}
%\nonumber \\
= m\sqrt{q_{2}^{2}} + q_{0\mu}q_{2}^{\mu} + \xi^{1}\Phi_{1}  - \frac{1}{2p_{2}^{2}(q_{0}q_{2})}\left(  {\frac{m}{\sqrt{q_{2}^{2}}}(q_{0}q_{2}) + q_{0}^{2} } \right)\Phi_{2}
\label{totalhamiltonian}
\end{eqnarray}
It is important to observe that though there are two primary first class constraints only one undetermined multiplier survives in the total Hamiltonian. This shows that effectively there is only one gauge degree of freedom. This feature distinguishes it from a genuine first order theory and has vital implications in the construction of the generator.
%\begin{eqnarray}
%q_{1}^{\mu}=x_{\mu}, \ q_{2}^{\mu}=\dot{x}^{\mu},\   p_{2}^{\mu}=-\frac{\partial{L}}{\partial\dot{x}^{\mu}},  \ p_{1}^{\mu}=-\frac{\partial{L}}{\partial\dot{x}^{\mu}}+\frac{d}{dt}\left( \frac{\partial{L}}{\partial{\ddot{x^{\mu}}}}\right) 
%\end{eqnarray}

 We now strongly impose the constraints (\ref{mPFC1}).  This is possible because the constraints (\ref{mPFC1}) merely eliminate the unphysical sector($q_{0\mu}$, $p_{0\mu}$) in favour of the physical variables. The constraint (\ref{mssc}) now read as
\begin{eqnarray}
%\nonumber
\omega_{1} =  p_{1}q_{2} +  m\sqrt{q_{2}^{2}} \approx 0;
\omega_{2} =  p_{1}p_{2} \approx 0
\label{mssc2}
\end{eqnarray}
The algebra of the remaining constraints can now be read off from (\ref{consalgebra}). We find
\begin{equation}
\left\lbrace \Phi_{2}, \omega_{2}\right\rbrace= 2p_{2}^{2}\left(p_{1}q_{2} \right); \ \ \ \ 
\left\lbrace \omega_{1}, \omega_{2}\right\rbrace = p_{1}^{2}-m^{2} 
\label{mconstalgb}
\end{equation}
If we take\\
\begin{equation}
\Phi_{2}^{\prime}=\left(p_{1}^{2}-m^{2} \right)\Phi_{2} - 2p_{2}^{2}\left( p_{1}q_{2}\right)\omega_{1}
\label{44}  
\end{equation}
then\\
\begin{equation}
\left\lbrace \Phi_{2}^{\prime}, \omega_{2}\right\rbrace=0
\end{equation}
If instead of the set of constraints $\Phi_{1}, \Phi_{2}, \omega_{1}, \omega_{2}$ we take their linear combinations in the form 
$\Phi_{1}, \Phi_{2}^{\prime}, \omega_{1}, \omega_{2}$ we find that only the PB between $ \omega_{1}$ and $\omega_{2}$ is non- involutive, given by the second equation in (\ref{mconstalgb}).
Clearly $\Phi_{1}, \Phi_{2}^{\prime}$ are first class and $\omega_{1}, \omega_{2}$ are second class.

 At this stage an important point is to be noticed. The canonical momentum  is set to be equal to the Lagrange multiplier $q_{0\mu}$ when we impose the constraint $\Phi_{1 \mu}$ strongly equal to zero. Thus $p_{1\mu}$ is completely arbitrary. It can be space, time or light-like. In the usual relativistic particle model $p_1^2 = m^2$ appears as a constraint of the theory. This is not the case here. In what follows we will assume that $p_1^2\ne m^2$. In other words we will consider a modified dispersion relation. 
%In the following thid modified dispersion relation will be shoen to be consistent from our analysis.
That such a modified dispersion relation is consistent will be demonstrated by our analysis. The case $p_1^2 = m^2$ will be seen as a singular point in our analysis and will be explored separately in section 3.2.

 Now the second class constraints $\omega_{1}, \omega_{2}$ may be strongly put equal to zero if the PBs $\{A, B\}$ are replaced by DBs $[A, B]$ {\footnote{Dirac brackets are denoted by $[$ $]$ to distinguish them from Poisson brackets which are written as $\left\lbrace  \right\rbrace$.}}.
 The non-vanishing DBs are given by {\footnote{Note that this computation is valid only when $p_1^2 \ne m^2$. The case $p_1^2 = m^2$ is thus a singular point in our analysis which is to be treated separately. }}
\begin{eqnarray}
\nonumber
\left[ {q_{1\mu}, q_{1\nu}}\right]
%&=&-\left\lbrace{q_{1\mu}, \omega_{1}} \right\rbrace\Delta_{12}^{-1}\left\lbrace {\omega_{2}, q_{1\nu}}\right\rbrace-\left\lbrace {q_{1\mu}, \omega_{2}}\right\rbrace\Delta_{21}^{-1}\left\lbrace {\omega_{1},q_{1\nu}}\right\rbrace
%\nonumber \\
%&=& \left( -q_{2\mu}\right)\left(-\frac{1}{p_{1}^{2}-m^{2}}\right)\left( -p_{2\nu}\right)-\left( p_{2\mu}\right)\left(\frac{1}{p_{1}^{2}-m^{2}}\right)\left( -q_{2\nu}\right)  
%\nonumber \\
=\frac{1}{p_{1}^{2}-m^{2}}\left({p_{2\mu}q_{2\nu}-q_{2\mu}p_{2\nu}} \right)     
\end{eqnarray}
%Now $\left\lbrace {q_{2\mu}, \omega_{1}}\right\rbrace=0, \left\lbrace {q_{2\mu}, \omega_{2}}\right\rbrace=p_{1\mu} $
%so\\
\begin{eqnarray}
\nonumber
\left[ {q_{1\mu}, q_{2\nu}}\right]
%&=&-\left\lbrace {q_{1\mu}, \omega_{1}}\right\rbrace\Delta_{12}^{-1}\left\lbrace{\omega_{2}, q_{2\nu}} \right\rbrace
%\nonumber \\
%&=&-\left( {q_{2\mu}}\right)\left({-\frac{1}{p_{1}^{2}-m^{2}}} \right)\left({-p_{1\nu}} \right)
%\nonumber \\
&=&-\frac{q_{2\mu}p_{1\nu}}{p_{1}^{2}-m^{2}}      
%\end{eqnarray}
%\begin{equation}
%\left\lbrace{p_{1\mu}, \omega_{1}} \right\rbrace=\left\lbrace {p_{1\mu}, \omega_{2}}\right\rbrace=\left\lbrace{p_{2\mu}, \omega_{2}} \right\rbrace=0   
%\end{equation}
%\begin{eqnarray}
%\nonumber
%\left\lbrace {p_{2\mu}, \omega_{1}}\right\rbrace&=&\left\lbrace{p_{2\mu}, \left({p_{1}q_{2}-m\sqrt{q_{2}^{2}}} \right) } \right\rbrace  
%\nonumber \\
%&=& -p_{1\mu}+\frac{m}{\sqrt{q_{2}^{2}}}q_{2\mu}
%\end{eqnarray}
%Then\\
%\begin{eqnarray}
\nonumber\\
\left[{q_{1\mu}, p_{2\nu}} \right]
%&=& -\left\lbrace{q_{1\mu}, \omega_{2}} \right\rbrace\Delta_{21}^{-1}\left\lbrace{\omega_{1}, p_{2\nu}} \right\rbrace   
%\nonumber \\
%&=&-\left( {p_{2\mu}}\right)\left({\frac{1}{p_{1}^{2}-m^{2}}} \right)\left({p_{1\nu}-\frac{m}{\sqrt{q_{2}^{2}}}q_{2\nu}} \right) 
%\nonumber \\
&=& \frac{1}{p_{1}^{2}-m^{2}}\left( {\frac{m}{\sqrt{q_{2}^{2}}}p_{2\mu}q_{2\nu} + p_{2\mu}p_{1\nu}}\right)
\label{basicdbs1}
 \end{eqnarray}
\begin{eqnarray}
%\nonumber
%\left[ {q_{2\mu}, p_{1\nu}}\right]&=& 0, \left[ {p_{1\mu},p_{1\nu}}\right]=0
%\nonumber \\
%\left[{p_{1\mu}, p_{2\nu}} \right]&=& 0, \left[ {p_{2\mu}, p_{2\nu}}\right]=0 
\nonumber \\
\left[{q_{1\mu}, p_{1\nu}} \right]&=& \eta_{\mu\nu}
% \left[{q_{2\mu}, q_{2\nu}} \right]=0     
\nonumber \\
\left[{q_{2\mu}, p_{2\nu}} \right]&=& \eta_{\mu\nu}-\frac{1}{p_{1}^{2}-m^{2}}\left({p_{1\mu}p_{1\nu} + \frac{m}{\sqrt{q_{2}^{2}}}p_{1\mu}q_{2\nu}} \right)
\label{50} 
\end{eqnarray}
The structure of the DBs is remarkable. We find that the coordinate algebra $\left[ {q_{1\mu}, q_{1\nu}}\right]$ becomes non -- commutative. Such non -- commutativity is generally known to modify the usual dispersion relation, both effects occuring at Planck scales. Our assumption of the condition $p_1^2 \ne m^2$ is thus consistent and it exhibits the appearance of coordinate noncommutativity in a simple setting including its connection with modified dispersion relation. As shown later in section 4, the treatment of the singular case $p_1^2 = m^2$ naturally leads to a vanishing algebra among the coordinates $q_{1\mu}$

  After the substitution of the basic PBs by the DBs there is some simplification of the constraint structure.
  The second class constraints $\omega_{1}$ and $\omega_{2}$ can now be strongly set equal to zero.
%($\omega_{1}$=$\omega_{2}$=0)
% which implies  that 
%\begin{eqnarray}
%\omega_1 = p_1q_2 +m \sqrt{q_2^2} = 0\nonumber\\
%\omega_2 = p_1p_2 = 0
%\label{constrntelimination}
%\end{eqnarray} 
The constraint $\Phi_{2}^{\prime}$(see equation (\ref{44}))  then reduces to,
\begin{eqnarray}
\nonumber
\Phi_{2}^{\prime}=\left(p_{1}^{2}-m^{2} \right)\Phi_{2}   
\end{eqnarray}
%We now have $\omega_{1} = 0$. Also $\left(p_{1}^{2}-m^{2} \right)$ is nonvanishing. Thus we can replace $\Phi_{2}^{\prime}$ again by $\Phi_{2}$. The set of remaining constraints is then taken as $\Phi_{1},\Phi_{2}$.
 Let us calculate the DB between the constraints.We find,
%\begin{eqnarray}
%\left[\Phi_{1} , \Phi_{2}^{\prime}\right] = -2\left[\left(p_1p_2\right)q_2^2\left(\left(p_1p_2\right) +\frac{m}{\sqrt{q_2^2}}\left(p_1q_2\right)\right) - p_2^2\left(p_1 q_2\right)\left(p_1q_2 +m \sqrt{q_2^2}\right) \right]
%\end{eqnarray}
\begin{eqnarray}
\left[\Phi_{1} , \Phi_{2}^{\prime}\right] = -2\left[\omega_2q_2^2\left(\omega_2 +\frac{m}{\sqrt{q_2^2}}\left(p_1q_2\right)\right) - p_2^2\left(p_1 q_2\right)\omega_1 \right] = 0
\end{eqnarray}
 since $\omega_1 = \omega_2 = 0$.

We now proceed towards the discussion on gauge symmetry. There are now two  first class constraints
$\Phi_{1}$, $\Phi_{2}^{\prime}$, both of which are primary.
%, $\left[ {\Phi_{1}, \Phi_{2}}\right]= 0 $ .
 So the generator of gauge transformation is \cite{D}
\begin{equation}
G=\epsilon^{1}\Phi_{1} + \epsilon^{2}\Phi^{\prime}_{2}
\label{generator}
\end{equation}
%
%$\Phi_{1}$ and $\Phi_{2}$ are first class 
%The canonical hamiltonian $H_{c}=-\omega_{1}=0$\\ 
%All the structure constants vanish.Both $\epsilon_{1}, \epsilon_{2}$ are independent gauge constants.\\
 Had it been a first order theory we would conclude that both $\epsilon^{1}, \epsilon^{2}$ are independent gauge parameters. However, due to the higher derivative nature we  have the additional  requirement 
\begin{eqnarray}
\frac{d}{d\tau}\delta{q_{1}^{\mu}} &=& \delta{q_{2}^{\mu}}\label{condition}
\end{eqnarray}
which follows from 
%(\ref{commutativity1}) and 
(\ref{varsgauge}).
Now
\begin{eqnarray}
\delta{q_{1}^{\mu}} = \left[q_{1}^{\mu}, G\right]
%&=& \epsilon_{1}\left[{q_{1}^{\mu}, \Phi_{1}}\right] + \epsilon_{2}\left[{q_{1}^{\mu}, \Phi_{2}} \right]\nonumber\\
 = 2\epsilon^{2}p_{2}^{2}q_{2}^{\mu}m\sqrt{q_{2}^{2}}
 \label{cmp1}
\end{eqnarray}
and
\begin{eqnarray}
\delta{q_{2}^{\mu}} = \left[q_{2}^{\mu}, G\right] = \epsilon^{1}q_{2}^{\mu}+2\epsilon^{2}q_{2}^{2}(p_{1}^{2} - m^{2})p_{2}^{\mu}
\label{55}
\end{eqnarray}
Hence using (\ref{condition})we get
%Here \\
%\begin{eqnarray}
%\nonumber
%\left[{q_{1}^{\mu}, \Phi_{1}}\right]=0
%\nonumber \\
%\left[{q_{1}^{\mu}, \Phi_{2}} \right]=-\frac{2p_{2}^{2}}{p_{1}^{2}-m^{2}}q_{2}^{\mu}m\sqrt{q_{2}^{2}}
%\end{eqnarray}
%$\delta{q_{2}^{\mu}} = \epsilon_{1}q_{2}^{\mu}+2\epsilon_{2}q_{2}^{2}p_{2}^{\mu}$
%now $\dot{q_{1}^{\mu}}=q_{2}^{\mu}$  so we require \\
\begin{eqnarray}
%\nonumber 
%\frac{d}{d\tau}\delta{q_{1}^{\mu}} &=& \delta{q_{2}^{\mu}}
%\nonumber \\
%or,
 \ \epsilon^{1}q_{2}^{\mu} + 2\epsilon^{2}q_{2}^{2}(p_{1}^{2} - m^{2})p_{2}^{\mu} = \frac{d}{d\tau}\left({2mp_{2}^{2}\sqrt{q_{2}^{2}}q_{2}^{\mu}\epsilon^{2}} \right)
\nonumber \\
%or, \ \epsilon_{1}q_{2}^{\mu}+ 2\epsilon_{2}q_{2}^{2}p_{2}^{\mu} &=& -\frac{2mp_{2}^{2}\sqrt{q_{2}^{2}}}{p_{1}^{2}-m^{2}}q_{2}^{\mu}\dot{\epsilon_{2}} - \epsilon_{2}\frac{d}{d\tau}\left({\frac{2mp_{2}^{2}\sqrt{q_{2}^{2}}}{p_{1}^{2}-m^{2}}q_{2}^{\mu}} \right) 
\end{eqnarray}
Taking scalar product with $q_{2\mu}$ and using $\Phi_{1} = p_{2}q_{2}\approx0$, we obtain\\
\\
\begin{eqnarray}
\epsilon^{1} = \frac{q_{2\mu}}{q_{2}^{2}}\frac{d}{d\tau}\left(2mp_{2}^{2}\sqrt{q_{2}^{2}}\epsilon^{2}q_{2}^{\mu} \right)
\label{gaugepara1}
\end{eqnarray}
Clearly, only one parameter $\epsilon^{2}$(say) is independent in the gauge generator G. This is also compatible with the observation that there is only one undetermined multiplier in the total Hamiltonian (\ref{totalhamiltonian}). The number of independent gauge parameters is thus shown to be less than the number of independent first class primary constraints.

	We next show that these findings are consistent with the reparametrization symmetry of the model to which the gauge symmetry is expected to  have a one to one correspondence. In the following we will show the mapping between the gauge parameter and the reparametrization parameter.\\

Consider the following reparametrization 
\begin{eqnarray}
\tau \to \tau + \Lambda
\label{repara}
\end{eqnarray}
where $\Lambda$ is an infinitesimal reparametrization parameter. By direct substitution we can verify that (\ref{repara}) is an invariance of (\ref{maction}). Now, under this reparametrization $x^{\mu}$ transforms as
\begin{equation}
x^{\prime\mu}\left(\tau\right) = x^{\mu}\left(\tau -\Lambda\right)
\end{equation}
The variation of $x^{\mu}$ is then
\begin{equation}
\delta{x^{\mu}} = x^{\prime\mu}(\tau) - x^{\mu}(\tau) =  -\Lambda\dot{x}^{\mu}\label{cmp3}
\end{equation}
%where $\Lambda$ is the reparametrization parameter.
From equation (\ref{cmp1}) we can write
\begin{equation}
\delta{x^{\mu}} = 2\epsilon^{2}p_{2}^{2}\dot{x}^{\mu}m\sqrt{q_{2}^{2}}
\label{cmp2}
\end{equation}
where we have used the identification(\ref{mnewcoordinate}). Comparing (\ref{cmp3}) and (\ref{cmp2}) we get the desired mapping
\begin{equation}
\Lambda =   - 2\epsilon^{2}p_{2}^{2}m\sqrt{q_{2}^{2}}
\label{reparavalue}
\end{equation}
Thus in our analysis an exact correspondence between the gauge and reparametrization symmetries is clearly demonstrated.
% parameter and the gauge parameter.
%\begin {appendix} 
%\renewcommand{\thesection}%{Appendix 
%\Alph{section}
%}			% redefine the command that creates the section heading.
%\setcounter{section}{0}										% redefine the command that creates the section no.

\subsection {Behaviour at the singular point ($p_1^2 = m^2$)} In the above Hamiltonian analysis of the relativistic particle model with curvature we have shown that $p_1^2$ is not constrained by the phase space structure and may be space , time or light-like. In our analysis we have assumed that $p_1^2$ is not equal to $m^2$. This is because the condition $p_1^2 = m^2$  is a singular point and must be treated separately. 
In the following we will discuss the construction of the gauge generator when the singular limit is assumed. This will also reveal the versatality of our method.

 The set of constraints after removal of  the unphysical sector $q_{0} = p_{1}$ is now given by,
 \begin{eqnarray}
\Phi_{1} = p_{2}q_{2} \approx 0;\Phi_{2} =p_{2}^{2}q_{2}^{2} + \alpha^{2} \approx 0;\omega_{1} = p_{1}q_{2} + m\sqrt{q_{2}^{2}} \approx 0 ; \omega_{2} = p_{1}p_{2} \approx  0\label{singularconstraints}
\end{eqnarray}
Here $\Phi_1$  and $\Phi_2$ are primary while $\omega_{1}$ and $\omega_{2}$ are secondary constraints. The constraint algebra shows that $\Phi_1$  and $\omega_{1}$ are first class while the rest are second class. So the constraint structure in the singular limit is different from that in the general case discussed above. Specifically the appearance of a secondary first class constraint is to be noted. 
%Also, the multiplier  $\xi_{2}$ in the the expreturns out to be vanishing. Hence again there is only one undetermined multiplier in 
The total Hamiltonian reads as
\begin{equation}
H_{T} = H_{c} + \xi^{1}\Phi_{1}
\label{singtotalH}
\end{equation}
where $H_{c}$ is the canonical Hamiltonian, given by 
\begin{equation}
H_{c} = m \sqrt{q_{2}^{2}} + p_{1}q_{2}
\end{equation}
Note that, as before, the total Hamiltonian consists of one undetermined multiplier
indicating one independent gauge degree of freedom.

    The construction of the  generator of the gauge transformations follow the course outlined in section 2. At first we strongly impose the second class  constraints $\left( {\Phi_{2}, \omega_{2}}\right)$ using the Dirac bracket formalism.
The non-trivial Dirac brackets among the phase-space variables are\\
\begin{eqnarray}
\nonumber \\
\left[ q_{1\mu}, q_{2\nu}\right] &=& \frac{q_{2}^{2}}{p_{2}^{2}} \frac{p_{2\mu}p_{2\nu}}{p_{1}q_{2}} 
\nonumber \\
\left[q_{1\mu}, p_{2\nu} \right] &=& - \frac{p_{2\mu}q_{2\nu}}{p_{1}q_{2}} 
\nonumber \\
\left[{q_{1\mu}, p_{1\nu}} \right]  &=& \eta_{\mu \nu}
\nonumber \\
\left[{q_{2\mu}, p_{2\nu}} \right] &=& \eta_{\mu \nu} - \frac{p_{1\mu} q_{2\nu}}{p_{1}q_{2}}
\nonumber \\
\left[ {q_{2\mu}, q_{2\nu}}\right] &=& \frac{q_{2}^{2}}{p_{2}^{2}} \frac{\left( {p_{1\mu}p_{2\nu} - p_{2\mu}p_{1\nu}}\right) }{p_{1}q_{2}}
\label{singDBs}
\end{eqnarray}
 Note that the coordinate algebra $\left[q_{1\mu}, q_{1\nu}\right]$ becomes commutative as a result of usual dispersion relation. This may be contrasted with the general case ($p_{1}^{2} \neq m^{2}$) leading to a noncommutative coordinate algebra (first equation in (\ref{basicdbs1})).
The  generator  is given by\\
\begin{equation}
G^{\prime} = \epsilon^{1} \Phi_{1} + \epsilon^{2}\omega_{1}
\label{ncgg}
\end{equation} 
Due to the presence of a secondary first class constraint($\omega_{1}$), $\epsilon^{1}$ and $\epsilon^{2}$ are not independent. There is a restriction \cite{BRR, BRR1}
\begin{eqnarray}
\frac{d\epsilon^2}{d\tau} - \sum_{a=1,2} \epsilon^a\left(V_a^{\ 2} + \xi^1C_{1 a}^{\ \ 2}\right)=0
\end{eqnarray} The other restriction(\ref{condition}) follows from the higher derivative nature . Interestingly, both the restrictions lead to the same condition 
\begin{equation}
\epsilon^{1} = \dot{\epsilon}^{2} + \xi^{1} \epsilon^{2}
\label{nc12}
\end{equation}
 To proceed further we require to express the Lagrange multiplier $\xi^1$ in terms of the phase space variables.  To this end we calculate $\dot{q}_2^\mu$ as
\begin{equation}
\dot{q}_2^\mu = \left[ {q}_2^\mu, H_T\right] 
\end{equation}
where $H_T$ is the total Hamiltonian given by equation (\ref{singtotalH}). Using the basic brackets (\ref{singDBs}) we get
%\begin{array}{rcl}
\begin{eqnarray}
%\nonumber
\dot{q}_2^\mu &=& \xi^1\left[{q}_2^\mu - \frac{p_1^\mu}{p_1^2 - m^2}\left(p_1q_2+m\sqrt{q_2^2}\right)\right] - \frac{q_2^2}{p_2^2p_1q_2}\left(\frac{m}{\sqrt{q_2^2}}(p_1q_2)+p_1^2\right)\nonumber\\
&&\left[p_2^\mu -\frac{p_1^\mu}{p_1^2 - m^2}\left(p_1p_2+\frac{m}{\sqrt{q_2^2}}p_2q_2\right)\right]
%\nonumber \\
\label{qdot}
\end{eqnarray}
%From equation (\ref{constrntelimination}) we find that $\omega_1 = p_1q_2 + m\sqrt{q_2^2} = 0$ and $\omega_2 = p_1p_2 = 0$. These constraints are strongly imposed. Also note the constraint algebra between the first class constraints $\Phi_1$ and $\Phi^{\prime}_2$ given by equation (\ref{53}). Since the bracket between these first class constraints vanishes strongly we can use $\Phi_1 = 0$ in (\ref{qdot}). Using these simplifications we get 
   \  \  Implementing the constraints (\ref{singularconstraints})  and simplifying, yields,
\begin{eqnarray}
\dot{q}_2^\mu = \xi^1{q}_2^\mu - \left(p_1^2 - m^2\right)\frac{q_2^2p_2^\mu}{p_2^2p_1q_2} ,
\end{eqnarray}
which immediately gives, on contraction with $q_{2\mu}$,
\begin{equation}
\xi^1 = \frac{1}{2q_2^2}\frac{d}{d\tau}\left(q_2^2\right)\label{ncvarxi1}
%\label{lagrange}
\end{equation}
\\

%\begin{equation}
%\xi^{1} = \frac{q_{2}\dot{q}_{2}}{q_{2}^{2}} 
%\label{ncvarxi1}
%\end{equation}\\
Using (\ref{nc12}, \ref{ncvarxi1}) we can express one of the gauge parameters appearing in $G^{\prime}$(\ref{ncgg}) in terms of the other. 
%The consistency of the scheme may again be checked by the method outlined above.
 We observe that in spite of the modified constraint structure corresponding to the singular point, the gauge generator may be consistently constructed by our general method. 
\subsection{A consistency check}
An important relation to check the consistency of our scheme is given by \cite{BRR, BRR1}
\begin{eqnarray}
\delta\xi^1 = \frac{d\epsilon^1}{dt} - \sum_{a= 1,2} \epsilon^a\left(V_a^{\  1} + \xi^1C_{1 a}^{\ \  1}\right)\label{parametervariation}
\end{eqnarray}
where the coefficients $V_a^{\  b}$ and $C_{ab}^{\ \ c}$ are defined by (\ref{2110}).
%\begin{eqnarray}
%\left[\Phi_a, H_C\right] = V_a^{\ b}\Phi_b ; \left[\Phi_a, \Phi_b\right] =C_{ab}^{\ \ c}\Phi_c
%\end{eqnarray}
Here both $V_a^{\ b}$ and $C_{ab}^{\ \ c}$ vanish.

   Now we can independently calculate $\delta\xi^1 $ for our theory and an exact agreement with (\ref{parametervariation}) will be demonstrated. 

	Taking the gauge variation on both sides of (\ref{ncvarxi1}) and substituting $\delta q_2^\mu$ from (\ref{55}) we arrive at,
\begin{eqnarray}
\delta\xi^1 = &-&\frac{1}{(q_2^2)^2}q_{2\mu}\left(\epsilon^1 q_2^\mu + 2\epsilon^2q_2^2(p_1^2-m^2)p_{2}^{\mu}\right)\frac{d}{d\tau}q_2^2\nonumber\\  &+& \frac{1}{q_2^2}\frac{d}{d\tau}\left[q_{2\mu}\left(\epsilon^1 q_2^\mu + 2\epsilon^2q_2^2(p_{1}^{2}-m^{2})p_{2}^{\mu}\right)\right]
\end{eqnarray}
Imposing the constraints (\ref{mPFC2}, \ref{mssc2}) and simplifying the ensuing algebra immediately reproduces(\ref{parametervariation}).This completes our consistency check.
%\begin{equation}
%\delta\xi_1 = \frac{d\epsilon_1}{d\tau}
%\end{equation}
%which is just the same as (\ref{parametervariation}), following from the general formalism.
\section{The rigid relativistic particle model}

	 The massless version of the model  known as `rigid relativistic particle', is obtained by setting $m=0$ in (\ref{maction}). It presents some unique features. We perform a detailed Hamiltonian analysis which is quite distinct from the earlier(massive) model due to a modified symplectic structure.  It is interesting to point out in this case that we will be able to find out an extra symmetry apart from the expected diffeomorphism  symmetry. This  is the $W_{3}$-symmetry \cite{RR}.
\subsection{Hamiltonian analysis}
The relativistic point particle theory with rigidity only has the action
%\footnote{contractions are abbreviated as $A^\mu B_\mu = AB$, $A^\mu A_\mu=A^2$} 
\begin{equation}
S=
%-m \int{\sqrt{\dot{x}^{2}}}d\tau+
\alpha\int{\frac{\left( \left( \dot{x}\ddot{x}\right)^{2}-\dot{x}^{2}\ddot{x}^{2} \right)^\frac{1}{2}}{\dot{x}^{2}}}d\tau\label{action}
\end{equation}
We introduce the new coordinates
\begin{equation}
q^{\mu}_1 = x^{\mu}\ \;\ \ q^{\mu}_2 = \dot{x}^{\mu}
\label{newcoordinate}
\end{equation}
The Lagrangian in these coordinates has a first order form given by
\begin{eqnarray}
% \nonumber
%L_{0} &=& -m\sqrt{q_{2}^{2}} + \alpha \frac{\left({\left({q_{2}\dot{q}_{2}} \right)^{2} - q_{2}^{2} \dot{q}_{2}^{2} } \right)^{\frac{1}{2}} }{q_{2}^{2}}
%\nonumber \\
L &=& \alpha \frac{\left({\left({q_{2}\dot{q}_{2}} \right)^{2} - q_{2}^{2} \dot{q}_{2}^{2} } \right)^{\frac{1}{2}} }{q_{2}^{2}} + q_{0}^{\mu}(\dot{q}_{1\mu} - q_{2\mu})
\label{independentfields}
\end{eqnarray}
where $q_{0}^{\mu}$ are the Lagrange multipliers that enforce the constraints
\begin{equation}
\dot{q}_{1\mu} - q_{2\mu} = 0
\label{27}
\end{equation}
Let $p_{0\mu}$, $p_{1\mu}$ and $p_{2\mu}$ be the canonical momenta conjugate to $q_{0\mu}$, $q_{1\mu}$ and $q_{2\mu}$ respectively. Then
\begin{eqnarray}
\nonumber
p_{0\mu} &=& \frac{\partial{L}}{\partial{\dot{q}_{0}^{\mu}}} = 0
\nonumber \\
p_{1\mu} &=& q_{0\mu}
\nonumber \\
p_{2\mu} &=& \frac{\alpha}{q_{2}^{2}\sqrt{g}} l_{\mu}
\label{regidmomenta}
\end{eqnarray}
 where,\\
\begin{eqnarray}
\nonumber
g &=& \left({\left({q_{2}\dot{q}_{2}} \right)^{2} - q_{2}^{2} \dot{q}_{2}^{2} } \right)
\nonumber \\
l_{\mu} &=& (q_{2}\dot{q}_{2})q_{2\mu} - q_{2}^{2}\dot{q}_{2\mu}
\end{eqnarray}
We immediately get the following
primary constraints
\begin{eqnarray}
\nonumber
\Phi_{0\mu} = p_{0\mu} \approx 0
\nonumber \\
\Phi_{1\mu} = p_{1\mu} - q_{0\mu} \approx 0
\label{PFC1}
\end{eqnarray}
and
\begin{eqnarray}
\nonumber
\Phi_{1} &=& p_{2}q_{2} \approx 0
\nonumber \\
\Phi_{2} &=& p_{2}^{2}q_{2}^{2} + \alpha^{2} \approx 0
\label{PFC2new}
\end{eqnarray}
The first set of constraints (\ref{PFC1}) are an outcome of our extension of the original Lagrangian. The canonical Hamiltonian following from the usual definition is given by
\begin{equation}
H_{C} =  q_{0\mu}q_{2}^{\mu}
 \label{CH}
\end{equation}
The total Hamiltonian is
\begin{equation}
H_{T} =  H_{C} + u_{0\mu}\Phi_{0}^{\mu} + u_{1\mu}\Phi_{1}^{\mu} + \lambda^{1}\Phi_{1} + \lambda^{2}\Phi_{2} 
\end{equation}
where $u_{0\mu}$, $u_{1\mu}$,$\lambda^{1}$ and $\lambda^{2}$ are as yet undetermined multipliers. Now, conserving the primary constraints we get
\begin{eqnarray}
\nonumber 
\dot{\Phi}_{0\mu} &=& \{\Phi_{0\mu}, H_T\}  = -q_{2\mu} + u_{1\mu} \approx 0
\nonumber \\
\dot{\Phi}_{1\mu} &=& \{\Phi_{1\mu}, H_T\}  = -u_{0\mu} \approx 0
\nonumber \\
\dot{\Phi}_{1} &=& \{\Phi_{1},  H_T\}  = -q_{0}q_{2}  \approx 0
\nonumber \\
\dot{\Phi}_{2} &=& \{\Phi_{2},  H_T\} = - 2(q_{0}p_{2})q_{2}^{2} \approx 0
\label{34}
\end{eqnarray}
We find that the multipliers  $u_{0\mu}$ and $u_{1\mu}$ are fixed
\begin{equation}
u_{0\mu} = 0 \ \ and \ \ u_{1\mu} = q_{2\mu}
\end{equation}
while new
secondary constraints are obtained
\begin{eqnarray}
\nonumber
\omega_{1} =  q_{0}q_{2}  \approx 0
\nonumber \\
\omega_{2} = q_{0}p_{2} \approx 0
\label{36new}
\end{eqnarray}
The last constraint in (\ref{34}) simplifies to $\omega_{2}$ since $q_{2}^{2} \neq 0$ as may be observed from the structure of the Lagrangian (\ref{independentfields}). The total Hamiltonian now becomes
\begin{equation}
H_{T} =  q_{0\mu}q_{2}^{\mu} + q_{2\mu}\Phi_{1}^{\mu} + \lambda^{1}\Phi_{1} + \lambda^{2}\Phi_{2} \label{tothamfin}
\end{equation}
	Preserving the constraints\ref{36new} we get $\dot{\omega}_{1} \approx 0$ ; $\dot{\omega}_{2} \approx -q_{0}^{2}$ thereby yielding a tertiary constraint $\Phi_{3} = q_{0}^{2} \approx 0$. This terminates the iterative process of obtaining constraints. The complete set of constraints is now given by,
\begin{eqnarray}
\Phi_{0\mu} &=& p_{0\mu} \approx 0\nonumber\\
\Phi_{1\mu} &=& p_{1\mu} - q_{0\mu} \approx 0\nonumber\\
\Phi_{1} &=& p_{2}q_{2} \approx 0\nonumber\\
\Phi_{2} &=& p_{2}^2q_{2}^2 + \alpha^2 \approx 0\nonumber\\
\omega_1 &=& q_0q_2\approx 0\nonumber\\
\omega_2 &=& q_0p_2\approx 0\nonumber\\
\Phi_3 &=& q_0^2 \approx 0
\end{eqnarray}
PBs among the constraints are given by \\
\begin{eqnarray}
\nonumber
\left\lbrace {\Phi_{0\mu}, \Phi_{1\nu}}\right\rbrace &=& \eta_{\mu\nu}
\nonumber \\
\nonumber
\left\lbrace {\Phi_{0\mu}, \omega_{1}}\right\rbrace &=& -q_{2\mu}
\nonumber \\
\left\lbrace {\Phi_{0\mu}, \omega_{2}}\right\rbrace &=& -p_{2\mu}
\nonumber \\
\left\lbrace {\Phi_{0\mu}, \Phi_{3}}\right\rbrace &=& -2q_{0\mu}
\nonumber \\
\left\lbrace {\Phi_{1}, \omega_{1}}\right\rbrace &=& -\omega_{1} =0
\nonumber  \\
\left\lbrace {\Phi_{1}, \omega_{2}}\right\rbrace &=& \omega_{2} = 0
\nonumber \\
\nonumber  \\
\left\lbrace {\Phi_{1}, \Phi_{2}}\right\rbrace&=& 0
\nonumber \\
\left\lbrace {\Phi_{2}, \omega_{1}}\right\rbrace &=& -2q_{2}^{2}\omega_2 = 0
\nonumber \\
\left\lbrace {\Phi_{2}, \omega_{2}}\right\rbrace &=& 2p_{2}^{2}( q_{0}q_{2} )=0
\nonumber \\
\left\lbrace{\omega_{1}, \omega_{2}} \right\rbrace &=& q_{0}^{2} =0
\label{algebra}       
\end{eqnarray}
Apparently, $\Phi_{0\mu}$, $\Phi_{1\mu}$, $\omega_1$, $\omega_2$ and $\Phi_3$ are second class. However we may substitute $\Phi_3$ by $\Phi_3^{\prime}$ where
\begin{equation} 
\Phi_3^{\prime} = \Phi_3 - 2q_{0}^{\nu} \Phi_{1\nu}\label{consnew}
\end{equation} 
One can easily verify that $\Phi_3^{\prime}$ commutes with all the constraints. The set of constraints are taken to be $\Phi_{0\mu}$, $\Phi_{1\mu}$, $\Phi_{1}$, $\Phi_{2}$, $\omega_1$, $\omega_2$ and $\Phi_3^{\prime}$ which are may be classified in  table 1.
\begin{table}[h]
%\begin{table}
\label{table:constraints}
\caption{Classification of Constraints}
\centering
\begin{tabular}{l  c  c}
\\[0.5ex]
\hline
\hline\\[-2ex]
& First class & Second class \\[0.5ex]
\hline\\[-2ex]
Primary &\ \ $\Phi_1, \Phi_2$ &\ \ $\Phi_{0\mu}, \Phi_{1\nu}$\\[0.5ex]
\hline\\[-2ex]
Secondary &\ $\Phi_3^{\prime}$ &\ \ $\omega_1, \omega_2$\\[0.5ex]
\hline
\hline
\end{tabular}
\end{table}
\subsection {Gauge symmetries and the emergence of the $W_3$ algebra}
 In the above we have considered the hamiltonian formulation of the  model (\ref{action}) in the first order approach. The expression of the total Hamiltonian (\ref{tothamfin}) suggests two independent gauge symmetries of the model. In the following we will construct the gauge generator and interpret the different gauge symmetries physically. 

 The variables $q_{0\mu}$ and their associated momenta $p_{0\mu}$ comprise the unphysical sector of the phase space. This is characterised by the second class pair $\Phi_{0\mu}$ and $\Phi_{1\mu}$. To find the gauge symmetries we have to eliminate these constraints. Considering their unphysical nature it will be appropriate to work in the reduced phase by putting
%In the following it will be convenient to work in the reduced phase space.
$\Phi_{0\mu}$ and $\Phi_{1\mu}$ equal to zero.
% and substituting the Poisson brackets (PB) by the corresponding Dirac Brackets (DB). It turns out that the DBs between the remaining phase space variables are equal to the corresponding PBs. 
The set of remaining constraints in the reduced phase space become 
\begin{eqnarray}
\Phi_{1} &=& p_{2}q_{2} \approx 0\nonumber\\
\Phi_{2} &=& p_{2}^2q_{2}^2 + \alpha^2 \approx 0\nonumber\\
\omega_1 &=& p_1q_2\approx 0\nonumber\\
\omega_2 &=& p_1p_2\approx 0\nonumber\\
\Phi_3 &=& p_1^2 \approx 0
\end{eqnarray}
Inspection of the algebra (\ref{algebra}) shows that in the reduced phase space
$ \Phi_{1}$, $\Phi_{2}$, $\omega_1$, $\omega_2$ and $\Phi_3^{\prime}$ form a first class set. In what follows it will be advantageous to rename the constraints as
\begin{eqnarray}
%\Phi_{0\mu} &=& p_{0\mu} \approx 0\nonumber\\
%\Phi_{1\mu} &=& p_{1\mu} - q_{0\mu} \approx 0\nonumber\\
\Omega_1 = \Phi_{1} &=& p_{2}q_{2} \approx 0\nonumber\\
\Omega_2 = \Phi_{2} &=& p_{2}^2q_{2}^2 + \alpha^2 \approx 0\nonumber\\
\Omega_3 = \omega_1 &=& p_1q_2\approx 0\nonumber\\
\Omega_4 = \omega_2 &=& p_1p_2\approx 0\nonumber\\
\Omega_5 = \Phi_3 &=& p_1^2 \approx 0
\end{eqnarray}
Also the canonical Hamiltonian is obtained from (\ref{CH}) as
\begin{equation}
H_c = \Omega_3\label{canham}
\end{equation}
The corresponding total Hamiltonian is
%The total Hamiltonian is
\begin{equation}
H_{T} =  H_{C}  + \lambda^{1}\Phi_{1} + \lambda^{2}\Phi_{2} 
\label{totham}
\end{equation}

%  As has been mentioned earlier we will follow the canonical algorithm of \cite{BRR} to construct the gauge generator. According to Dirac's conjecture the gauge generator $G$ is given by
%\begin{equation}
%G = \epsilon^a\Omega_a\label{gauge}
%\end{equation}
%Note that all first class constraints appear in the expression multiplied by the corresponding gauge parameters $\epsilon^a$. However not all of these parameters are independent. For the usual first order lagrangean theories the number of independent gauge parameters is equal to the number of primary first class constraints \cite{BRR}. The dependent gauge parameters are expressed in terms of the independent ones from the following set of equations:
%\begin{equation}
%  0 = \frac{d\epsilon^{a_2}}{dt}
% -\epsilon^{a}\left(V_{a}^{a_2}
%+\lambda^{b_1}C_{b_1a}^{a_2}\right)
%\label{219}
%\end{equation}
%Here the coefficients $V_{a}^{a_{1}}$ and $C_{b_1a}^{a_1}$ are the structure
%functions of the involutive algebra, defined as
%\begin{eqnarray}
%\{H_c,\Omega_{a}\} = V_{a}^b\Omega_{b}\nonumber\\
%\{\Omega_{a},\Omega_{b}\} = C_{ab}^{c}\Omega_{c}
%\label{2110}
%\end{eqnarray}
%Solving (\ref{219}) it is possible to choose $a_1$ independent
%gauge parameters from the set $\epsilon^{a}$ and express $G$ of
%(\ref{gauge}) entirely in terms of them. The relations (\ref{219}) are obtained 
%from the commutativity of the gauge transformation.
 
 As done previously, the gauge generator is written as a combination of all the first class constraints,
\begin{equation}
G = \sum_{a=1}^{5} \epsilon^{a}\Omega_{a}.
\label{rigidsumG}
\end{equation} 
 However,due to the presence of secondary first-class constraints, the parameters of gauge transformation($\epsilon_{a}$) are not independent. The independent parameters will be isolated by using \ref{219}. The first step in this direction is to calculate the structure functions $V_{a}^b$ and $C_{ab}^{c}$ using their definitions (\ref{2110}). A straightforward calculation gives the following nonzero values:
\begin{eqnarray}
V_1^{\ 3} &=& 1,\hspace{1cm} V_2^{\ 4} = 2q_2^{\ 2}\nonumber\\
V_4^{\ 5} &=& 1
 \label{v's}
\end{eqnarray}
and
\begin{eqnarray}
C_{13}^{\  3} &=& -1 ,\hspace{1cm} C_{14}^{\  4} = 1 \nonumber\\
C_{23}^{\   4} &=& -2q_2^{\ 2},\hspace{1cm} C_{24}^{\  3} = 2p_2^{ 2}\nonumber\\
C_{34}^{\ \ 5} &=& 1 
\label{c's}
\end{eqnarray}

Using these(\ref{v's}, \ref{c's})  in the master equation (\ref{219}) we arrive at the following equations
\begin{eqnarray}
{\dot{\epsilon}}^3 - \epsilon^1 + \lambda^1\epsilon^3 - 2\lambda^2\epsilon^4p_2^{\ 2} = 0\nonumber\\
{\dot{\epsilon}}^4 - 2q_2^{\ 2}\epsilon^2 - \lambda^1\epsilon^4 + 2\lambda^2\epsilon^3q_2^{\ 2} = 0\nonumber\\
{\dot{\epsilon}}^5 = \epsilon^4\label{gaugerelations}
\end{eqnarray}
Certain points are immediately apparent from the above equations. Out of the five gauge parameters $\epsilon^a$ three parameters may be expressed in terms of the remaining two using the equations(\ref{gaugerelations}). It is most convenient to take $\epsilon^3$
and $\epsilon^5$ as independent. 

To proceed further we require to work out the Lagrange multipliers in (\ref{totham}). For that purpose we first compute $\dot{q}_{2} = \left\lbrace {q_{2}, H_{T}}\right\rbrace $. Using the expression for total Hamiltonian given in(\ref{totham})we get 
   \begin{equation}
\dot{q}_{2} = \lambda^{1}q_{2\mu} + 2\lambda^{2}q_{2}^{2}p_{2\mu}
\label{rigidq2dot}
\end{equation}  
  Now scalar multiplying the above equation(\ref{rigidq2dot}) by $q_{2}^{\mu}$ and $p_{2}^{\mu}$ we obtain respectively solutions for $\lambda^1$ and $\lambda^2$ as
\begin{eqnarray}
\lambda^1 &=& \frac{q_2{\dot{q}}_2}{q_2^{\ 2}}\nonumber\\
\lambda^2 &=& 
\frac{\alpha
\sqrt{g}}
{2p_2^{2}q_2^{\ 4}}
\end{eqnarray}
Substituting these results for $\lambda^{1}$, $\lambda^{2}$ and $\epsilon^{4}=\dot{\epsilon}^{5}$ in the first two equations of (\ref{gaugerelations}) one can express 
\begin{eqnarray}
 \epsilon^1 &=& {\dot{\epsilon}}^3  + \frac{q_2{\dot{q}}_2}{q_2^{\ 2}}\epsilon^3 - \frac{\alpha\sqrt{g}}{q_2^{\ 4}}\dot{\epsilon}^{5} \nonumber\\
\epsilon^2 &=& \frac{1}{2q_2^{\ 2}}
\left({\ddot{\epsilon}}^5 - \frac{q_2{\dot{q}}_2}{q_2^{\ 2}}{\dot{\epsilon^5}} + \frac{\alpha\sqrt{g}}{p_2^{\ 2}q_2^{\ 2}}\epsilon^3\right)% \nonumber\\
%\epsilon^4 &=& {\dot{\epsilon}}^5 
 \label{gaugerelations1}
\end{eqnarray}
Note that the theory under investigation is a higher derivative theory. So the gauge transformations are additionally subject to the condition (\ref{varsgauge}). After simplification this condition reduces to
\begin{eqnarray}
2q_2{\dot{p}}_1\epsilon^5 = \left(q_2{\dot{p}}_2 + \frac{\alpha
\sqrt{g}}
{q_2^{\ 2}}\right)
\end{eqnarray}
Using the constraints of the theory this condition reduces to a tautological statement $0 = 0$. So, for this particular model, (\ref{varsgauge}) does not impose any new conditions on the gauge parameters. We thus find that there are two independent gauge parameters in the expression of the gauge generator $G$.Taking $\epsilon^{3}$ and $\epsilon^{5}$ to be independent, the expression for the gauge generator becomes(\ref{rigidsumG})
\begin{eqnarray}
G &=& \left( {\dot{\epsilon}}^3  + \frac{q_2{\dot{q}}_2}{q_2^{\ 2}}\epsilon^3 - \frac{\alpha\sqrt{g}}{q_2^{\ 4}}\dot{\epsilon}^{5}\right) \Omega_{1}+ \nonumber\\
&&  \frac{1}{2q_2^{\ 2}}
\left({\ddot{\epsilon}}^5 - \frac{q_2{\dot{q}}_2}{q_2^{\ 2}}{\dot{\epsilon^5}} + \frac{\alpha\sqrt{g}}{p_2^{\ 2}q_2^{\ 2}}\epsilon^3\right)\Omega_{2}+\epsilon^{3}\Omega_{3}+\dot{\epsilon}^{5}\Omega_{4}+\epsilon^{5}\Omega_{5}
\end{eqnarray}
that there are only two independent parameters  is consistent with the fact that there were two independent Lagrange multipliers in the expression of the total hamiltonian (\ref{totham}). Note, however, the distinction from the massive model discussed in the previous section. There we found only one independent gauge degree of freedom which was shown to have a one to one correspondence with the diffeomorhism invariance of the model. Clearly, the rigid relativistic particle is endowed with more general symmetries as is indicated by its gauge generator. 
	
	In order to unravel the meaning of the additional gauge symmetry we calculate the gauge variations of the dynamical variables, defined as $\delta{q} = \left\lbrace {q, G}\right\rbrace $. These are given by,
\begin{eqnarray}
\nonumber
\delta{q_{1}^{\mu}} &=& \epsilon^{3}q_{2}^{\mu}+ \dot{\epsilon}^{5}p_{2}^{\mu} + 2\epsilon^{5}p_{1}^{\mu}\\
\nonumber
\delta{q_{2}^{\mu}} &=& \left( {{\dot{\epsilon}}^3  + \frac{q_2{\dot{q}}_2}{q_2^{\ 2}}\epsilon^3 - \frac{\alpha\sqrt{g}}{q_2^{\ 4}}\dot{\epsilon}^5 } \right)  q_{2}^{\mu} + \left({\ddot{\epsilon}}^5 - \frac{q_2{\dot{q}}_2}{q_2^{\ 2}} {\dot{\epsilon^5}} + \frac{\alpha\sqrt{g}} {p_2^{\ 2}q_2^{\ 2}}\epsilon^3\right)p_{2}^{\mu} + \dot{\epsilon}^{5}p_{1}^{\mu}\\
\nonumber
\delta{p_{1}^{ \mu}} &=& 0\\
\nonumber
 \delta{p_{2}^{\mu}} &=& -\left( {{\dot{\epsilon}}^3  + \frac{q_2{\dot{q}}_2}{q_2^{\ 2}}\epsilon^3 - \frac{\alpha\sqrt{g}}{q_2^{\ 4}}\dot{\epsilon}^5} \right)  p_{2}^{\mu} \nonumber\\
 &&
 -\frac{p_{2}^{2}} {q_{2}^{2}} \left({\ddot{\epsilon}}^5 - \frac{q_2{\dot{q}}_2}{q_2^{\ 2}} {\dot{\epsilon^5}} + \frac{\alpha\sqrt{g}} {p_2^{\ 2}q_2^{\ 2}}\epsilon^3\right)q_{2}^{\mu}  -\epsilon^{3}p_{1}^{\mu}
%\nonumber
\label{dwinvariance}
\end{eqnarray}
%Let us consider that $\epsilon^{3}$ corresponds to reparametrization parameter and $\epsilon^{5}$ corresponds to w-symmetry which we shall show in the next section. Justification about $\epsilon^{3}$ is given bellow.
	
%Let us consider a reparametrization $\tau \rightarrow \tau + \Lambda $, where $\Lambda$ is an infinitesimal reparamtrization parameter then 
%	\begin{equation}
%\delta{x^{\mu}} = x^{\mu}(\tau - \Lambda) -  x^{\mu}(\tau ) = - \Lambda\dot{x}^{\mu}\label{repinva}
%\end{equation}
%Using our definition (\ref{newcoordinate}) this becomes
%\begin{equation}
%\delta{q_1^{\ \mu}} = - \Lambda q_2^{\ \mu}\label{repinv1}
%\end{equation}
\begin{figure}[th]
\centering
\includegraphics{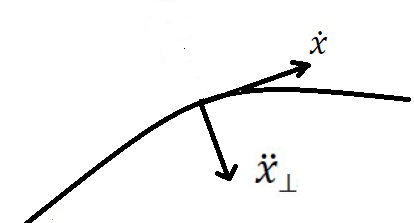}
\caption{	orthogonal frame attached with particle world trajectory}
\label{fig:figure1}
\end{figure}
Now it would be customary to identify the geometrical origin of these independent gauge transformations. It is well known  that  the most general deformation of the rigid particle trajectory may be resolved as
%%by an orthogonal coordinate system written as
\cite{RR0}
\begin{equation}
\delta_{\beta, \eta}{x} = \beta(t)\dot{x} + \eta(t)\ddot{x}_{\perp}
\label{deformation}
\end{equation}  
where $\ddot{x}_{\perp}^{\mu} = \frac{l^{\mu}}{\dot{x}^{2}}$ is orthogonal to the tangent space mapped by $\dot{x}$(see figure.$1$). The coefficients $\beta$, $\eta $ are 
%some terms containing 
respectively the diffeo-morphism and w-morphism parameters. Now we rewrite the first equation of (\ref{dwinvariance}) as
\begin{equation}
\delta{q_{1}^{\mu}} = \epsilon^{3}{\dot{x}}^{\mu} + \dot{\epsilon}^{5}\frac{\alpha}{\sqrt{g}}\ddot{x}^{\mu}_{\perp} + 2\epsilon^{5}p_{1}^{\mu}
\label{dwvariance1}
\end{equation}
As we have noted earlier, the gauge variations (\ref{dwinvariance}) contain two independent gauge parameters $\epsilon^{3}$ and $\epsilon^{5}$. Let us first assume 
\begin{equation}
\epsilon^{3} \ne  0\ \ \  {\rm{and}}  \ \ \ \epsilon^{5} = 0
\label{D}
\end{equation}
 Substituting this condition in  (\ref{dwvariance1}) we get 
\begin{equation}
\delta{q_{1}^{\mu}} = \epsilon^{3}{\dot{x}}^{\mu} 
\end{equation}
Comparing the above with (\ref{deformation}) we can easily see that the gauge generator subject to the limiting condition (\ref{D}) generates diffeomorphism(with $\epsilon^{3}=\beta$) and the corresponding gauge symmetry is identified with diffeomorphism invariance.
% which is reconfirmed from (\ref{cmp3}).
  This identification may be confirmed by working out the variations of other phase space variables under reparametrization and comparing them with the gauge variation generated by $G$, subject to the same condition (\ref{D}).

       It is now clear that the gauge generator subject to the other extreme condition
\begin{equation}
\epsilon^{3} =  0\ \ \  {\rm{and}}  \ \ \ \epsilon^{5} \ne 0 
\label{W}
\end{equation}
 generates symmetry transformations 
other than diffeos. Substituting (\ref{W}) in (\ref{dwvariance1}) we now get
\begin{equation}
\delta{q_{1}^{\mu}} =  \dot{\epsilon}^{5}\frac{\alpha}{\sqrt{g}}\ddot{x}^{\mu}_{\perp} + 2\epsilon^{5}p_{1}^{\mu}
\label{dwvariance12}
\end{equation}
Looking back at the second equation of (\ref{regidmomenta})  we find that the linear term containing $p_{1}^{\mu}$ in the variation (\ref{dwvariance12}) can be neglected, as $p_{1}^{\mu}$ is a Lagrange multiplier. Comparing with (\ref{deformation}) we can then associate (\ref{dwvariance12}) with w - transformation, by identifying $ \eta = \dot{\epsilon}^{5}\frac{\alpha}{\sqrt{g}}$ . Thus the additional symmetry corresponding to (\ref{W}) is nothing but
 W - symmetry. It is now straightforward to show that the two different symmetries together satisfy the ${\rm{W}}_3$ algebra.
% again comparing (\ref{deformation}) where the linear term containing $p_{1}^{\mu}$ can be neglected from its own definition(second equation of (\ref{regidmomenta}) ) as it is a Lagrange multiplier, so arbitrary. We again can reconfirm the identification of $\epsilon^{5}$ as the gauge parameter for w-symmetry as they obeys the following algebra \cite{RR0, RR}. 
 Let us denote the  transformations of category 1 (diffeomorphisms) by the superscript `$D$' and the category 2  transformations by `$W$'. Detailed calculations on all the phase-space variables show that
 \begin{eqnarray}
 \nonumber
\left\lbrace { \delta^{(D)}_{\epsilon^{3}_{1}} , \delta^{(D)}_{\epsilon^{3}_{2}} }\right\rbrace  &=& \delta^{(D)}_{\epsilon^{3}}; \ \ \ with  \ \ \ \ \epsilon^{3}= \dot{\epsilon}_{1}^{3} \epsilon_{2}^{3} - \epsilon_{1}^{3} \dot{\epsilon}_{2}^{3}  \\
 \nonumber
\left\lbrace { \delta^{(D)}_{\epsilon^{3}} , \delta^{(W)}_{\epsilon^{5}}}\right\rbrace  &=& \delta^{(W)}_{\epsilon^{\prime 5}}; \ \ \ with  \ \ \ \ \epsilon^{\prime5}= -\epsilon^{3}\epsilon^{5}\\ 
 \nonumber
\left\lbrace { \delta^{(W)}_{\epsilon^{5}_{1}} , \delta^{(W)}_{\epsilon^{5}_{2}} }\right\rbrace  &=& \nonumber \delta^{(W)}_{\epsilon^{5}}; \ \ \ with  \ \ \ \ \epsilon^{5}= \frac{p_{2}^{2}}{q_{2}^{2}}\left( \dot{\epsilon}_{2}^{5} \epsilon_{1}^{5} - \epsilon_{2}^{5} \dot{\epsilon}_{1}^{5}  \right) 
\end{eqnarray}
%This is  
which is nothing but the  $W_3$ algebra. 
%So the category 2 transformations correspond to the $W$ -- symmetry.

%\end{appendix}

\section{Conclusion}

 The issues of gauge symmetry, their classification, categorization etc. have been the subject of intense research over a long period of time. Dirac's constrained Hamiltonian approach \cite{D} provides a powerful tool for the analysis of gauge invariances and there are different algorithms available in the literature \cite{C, BRR, BRR1} for constructing the gauge generator. A remarkable feature is the equality of the  number of independent gauge parameters with the number of independent primary first class constraints ( PFC). This has been elaborated in \cite{HTZ, BRR, BRR1} where a compact Hamiltonian algorithm is also provided to obtain the most general gauge generator. All these works refer, however, to  theories whose Lagrangians are functions of coordinates and velocities  ( i.e. first order theories ). Now there is no apparent reason why one should be limited to  first order theories only. In fact higher derivative theories have many welcome features \cite{1b, 2b, 3b, 4b, BOS, Sot} and are being studied over a long period of time \cite{lw,gitman1, gitman2,P, N, P, DN, HH, R, S, K, cl, AGMM, M}. Specifically, their relevance is apparent in the context of modified theories of gravity which find application in quantum gravity \cite{BOS} and cosmology \cite{Sot}. Though there are a large number of works on the Hamiltonian formulations of the higher derivative theories \cite{Ostro, BGPR, pl1,  N, B, Mo, DS, gitman2, AGMM}, the problem of gauge symmetry in such theories is scarcely highlighted. In fact there are peculiar surprises in this context. Thus in the relativistic particle model with curvature \cite{P, pl1, N}, which is a higher derivative theory, we find that there are two  primary first class constraints. However, it is expected that this theory should have one independent gauge symmetry corresponding to the invariance of the model in the reparametrization of its trajectory in space-time. Also there should exist an exact mapping between the two sets of parameters, as has been demonstrated in other contexts \cite{BMS1, BMS2, GGS, MS}. 
%Thus there is a violation of Dirac conjecture which remains unnoticed and therefore unexplained. 

              In the present paper we have addressed the problem of gauge symmetry in higher derivative theories. We have discussed a Hamiltonian method of abstracting the independent gauge invariances for the higher order Lagrangians. The new features that emerged in our analysis pinpoint the reason for the mismatch between the  number of independent gauge parameters with the number of  primary first class constraints ( PFC). This is shown to be due to the fact that the transformation of the tangent bundle to coordinate manifold to the cotangent 
bundle  (phase space) is performed by adding Lagrange
multipliers  enforcing the proper relation between new  coordinates  and time derivatives of the original 
ones. These restrictions may impose further conditions between the gauge parameters so that the number of independent gauge parameters in the gauge generator may become less than the number of independent PFCs.

After discussing the general formalism we have applied it to different particle models containing higher derivative terms. Taking the example of the relativistic particle with curvature \cite{P, pl1} we have applied our Hamiltonian method to construct the gauge generator. The latter is shown to consist of one independent parameter, thereby resolving the paradox mentioned earlier. Also we have devised the exact mapping between this gauge parameter and the reparametrization parameter. This is a welcome result in view of the equivalence in the gauge and reparametrization parameters of generally covariant models \cite{BMS1, BMS2, GGS, MS}.
	
	As a byproduct of our analysis a close connection between noncommutativity and modified dispersion relations was revealed. Normally one introduces by hand a noncommuting algebra which leads to a modified dispersion relation. Here we observe that taking $p_{1}^{2} \neq m^{2}$ naturally yields a nonvanishing algebra among the coordinates while a separate treatment of $p_{1}^{2} = m^{2}$ yields the standard algebra.

         The next example was the rigid relativistic particle model. Though it is the massless limit of the previous example, the phase space structure is found to be very different.We have provided a complete Hamiltonian analysis of the model. Specifically, we have constructed the Hamiltinian gauge generator. Using this generator, the variations of the  phase space variables were computed. The gauge generator was found to contain two independent gauge parameters. Restoring to a geometric approach we have identified these symmetries with the diffeomorphism and W-morphisms and explicitly demonstrated the $W_{3}$-algebra. Note that previously the symmetries was demonstrated on-shell\cite{RR} and a consistent Hamiltonian analysis was lacking.  By casting the equations of motion of this model in the Bossinesq form it was earlier demonstrated that the model is endowed with larger a symmetry group. It is interesting that our Hamiltonian method straightforwardly exhibits the full symmetry group and the $W_3$ -- algebra is established in the Hamiltonian approach without taking recourse of the equations of motion.

       To conclude, we have provided a general method of analysing the gauge symmetries of higher derivative Lagrangian theories with a complete algorithm. This was applied here for various relativistic particle models. 
%As we have mentioned in the introduction, many variants of the relativistic particle models have been studied in the literature. 
These models show remarkable variety of Hamiltonian structures and gauge invariances. However our general method is demonstrated to be fully capable of unravelling the gauge invariances of the different models. 
%More detailed analysis will be given in a future communication. 
It appears that the method presented here will be useful as a  general algorithm to treat the gauge symmetry of higher derivative theories. This is not a mean achievment considering the importance of the higher derivative theories in modern theoretical physics including various theories of gravity \cite{1b, 2b, 3b, 4b, BOS, Sot}.

\section*{Acknowledgement}
One of the authors(PM) likes to acknowledge the hospitality of S. N. Bose National Centre For Basic Sciences during his stay as visiting associate when much of the work was done. B.P. thanks the Council for Scientific and Industrial Research, Govt. of India, for financial support. 
 %%%%%%%%%%%%%%%%%%%%%%%%%%%%%%%%%%%%%%

%\end{thebibliography} 
\end{document}